\documentclass[12pt]{article}

\usepackage[a4paper,left=30mm,right=20mm,top=20mm,bottom=20mm]{geometry}
\usepackage{graphics}
\usepackage{amsmath}
\usepackage{epsfig}
\usepackage{cite}
\usepackage{xcolor}

\newcommand{\z}{&&\hspace*{-1cm}}

\newcommand{\bea}{\begin{eqnarray}}
\newcommand{\eea}{\end{eqnarray}}
\newcommand{\be}{\begin{equation}}
\newcommand{\ee}{\end{equation}}

\newcommand{\ar}{a_s}

\title{QCD analytic coupling}

\author{I.A.~Zemlyakov$^{1,2}$, I.L. Chuev$^{3}$, A.V.~Kotikov$^{4}$}
\date{\vspace{-5ex}}
\begin{document}

\maketitle

\begin{center}

  {\it $^1$Department of Physics, Universidad Tecnica Federico Santa Maria,\\
  Avenida Espana 1680, Valparaiso, Chile,}\\
  {\it $^2$Tomsk State University, 634010 Tomsk, Russia}\\
  {\it $^3$Department of Theoretical Physics,
  Moscow Institute for Physics and Technology,
  141701 Dolgoprudny, Russia}\\
  {\it $^4$Bogoliubov Laboratory of Theoretical Physics,
  Joint Institute for Nuclear Research, 141980 Dubna, Russia}\\

\end{center}


\begin{center}

{\bf Abstract }

\end{center}

A brief overview of QCD analytic coupling is presented, mainly following the results obtained in
\cite{Kotikov:2022sos}.
An application to the pion-photon transition form factor is demonstrated.

\noindent

\label{sec:intro}
\section{Introduction}
According to the general principles of (local) quantum field theory (QFT) \cite{Bogolyubov:1959bfo},
observables in a spacelike region (i.e., in Euclidean space)
\footnote{Here we consider only observables in the spacelike region. The analytic coupling in the timelike region can be found in Ref. \cite{KoZe23}.}
can have singularities only for negative values of their argument $Q^2$.
However, for large $Q^2$ values, these observables are usually represented as power expansions in the strong
coupling $\alpha_s(Q^2)$,
which has a ghostly singularity, the so-called Landau pole, at $Q^2 = \Lambda^2$. Therefore, to restore the analyticity of the considered expansions,
this pole in the strong coupling should be removed.

The strong coupling $\alpha_s(Q^2)$ obeys the renormalization group equation
\be
L\equiv \ln\frac{Q^2}{\Lambda^2} = \int^{\overline{a}_s(Q^2)} \, \frac{da}{\beta(a)},~~ \overline{a}_s(Q^2)=\frac{\alpha_s(Q^2)}{4\pi}\,
\label{RenGro}
\ee
with some boundary condition and the QCD $\beta$-function:
\be
\beta(\ar) ~=~ -\sum_{i=0} \beta_i \overline{a}_s^{i+2}
=-\beta_0 \overline{a}_s^{2} \, \Bigl(1+\sum_{i=1} b_i \ar^i \Bigr),~~ b_i=\frac{\beta_i}{\beta_0^{i+1}}\,, ~~
\ar(Q^2)=
\beta_0\,\overline{a}_s(Q^2)\,,
\label{beta}
\ee
where
\be
\beta_0=11-\frac{2f}{3},~~\beta_1=102-\frac{38f}{3},~~
\label{beta_i}
\ee
for $f$ active quark flavors. Currently, the first five coefficients, i.e., $\beta_i$ with $i\leq 4$, are known exactly \cite{Baikov:2016tgj}.

Note that in Eq.~(\ref{beta})
we have incorporated the first coefficient of the QCD $\beta$-function into the $\ar$-definition, as is usually done in the context of
analytic coupling (see, e.g., Refs.~\cite{ShS,MSS,BMS1,Bakulev:2006ex}).

Thus, already at leading order (LO), where $\ar(Q^2)=\ar^{(1)}(Q^2)$, we obtain from Eq.~(\ref{RenGro})
\be
\ar^{(1)}(Q^2) = \frac{1}{L}\,,
\label{asLO}
\ee
i.e., $\ar^{(1)}(Q^2)$ indeed contains a pole at $Q^2=\Lambda^2$.

In Refs.~\cite{ShS,MSS}, an efficient approach was developed to eliminate the Landau singularity without introducing extraneous infrared regulators,
such as the gluon effective mass.
%
This method is based on a dispersion relation that relates the new analytic coupling $A_{\rm MA}(Q^2)$ to the spectral function $r_{\rm pt}(s)$
obtained in the framework of perturbation theory (PT).
At LO this gives
    \be
A^{(1)}_{\rm MA}(Q^2)
= \frac{1}{\pi} \int_{0}^{+\infty} \,
\frac{ d s }{(s + t)} \, r^{(1)}_{\rm pt}(s),~~ r^{(1)}_{\rm pt}(s)= {\rm Im} \; a_s^{(1)}(-s - i \epsilon) \,.
\label{disp_MA_LO}
\ee
The approach of \cite{ShS,MSS} follows the corresponding results \cite{Bogolyubov:1959vck} obtained in the framework of Quantum Electrodynamics.

So, we repeat once again: the spectral function in the dispersion relation (\ref{disp_MA_LO})
is taken from PT directly,
and the analytic coupling $A_{\rm MA}(Q^2)$
is restored using the
dispersion relation. This approach is usually
called the {\it Minimal Approach} (MA) (see, e.g., \cite{Cvetic:2008bn})
or the {\it Analytical Perturbation Theory} (APT) \cite{ShS,MSS}.
\footnote{An overview of other similar approaches can be found in \cite{Bakulev:2008td}, including
  \cite{Nesterenko:2003xb} which are close to APT.}

Thus, MA QCD is a very convenient approach that combines the analytic properties of QFT quantities and the results
obtained in the framework of perturbative QCD, leading to the appearance of the MA coupling $A_{\rm MA}(Q^2)$,
which is close to the usual
strong coupling $a_s(Q^2)$ in the limit of large $Q^2$ values and completely different from $a_s(Q^2)$ for small $Q^2$ values,
i.e., for $Q^2 \sim \Lambda^2$.

A further development of APT is the so-called fractional APT (FAPT) \cite{BMS1,Bakulev:2006ex},
which extends the construction principles
described above to PT series starting from non-integer  coupling-powers.
In the QFT framework, such series arise for quantities that have
non-zero anomalous dimensions.
Compact expressions for quantities within the FAPT framework were obtained mainly at LO, but this approach was also used in higher orders
(see, \cite{BMS1,Bakulev:2006ex}),
mainly by re-expanding the corresponding couplings in powers of the LO coupling, as well as using some approximations.

Here we give an overview of the main properties of
MA couplings in the FAPT framework, obtained in Ref.~\cite{Kotikov:2022sos}
using the so-called $1/L$-expansion. Note that for an ordinary coupling, this expansion is applicable only for large $Q^2$ values, i.e., for $Q^2>>\Lambda^2$.
However, as shown in \cite{Kotikov:2022sos}, the situation is quite different in the case of the analytic coupling,
and this $1/L$-expansion is applicable for all $Q^2$ values.
This is due to the fact that the non-leading expansion corrections vanish not only
at $Q^2 \to \infty$, but also at $Q^2 \to 0$,
\footnote{The absence of high-order corrections for $Q^2 \to 0$ was  discussed also in Refs.~\cite{ShS,MSS}.}
which leads only to nonzero (small) corrections in the region $Q^2 \sim \Lambda^2$. 

Below we consider the forms of
the MA couplings and their (fractional) derivatives obtained in \cite{Kotikov:2022sos} and valid in
principle in any PT order. However, in order to avoid cumbersome formulas, but at the same time to show the main features of the approach
obtained in \cite{Kotikov:2022sos}, we confine ourselves to considering only the first two PT orders.

\section{Strong coupling}
\label{strong}

As shown in the Introduction, the strong coupling $a_s(Q^2)$ obeys the renormalization group equation (\ref{RenGro}).
When $Q^2>>\Lambda^2$, Eq.~(\ref{RenGro}) can be solved by iterations in the form of a $1/L$-expansion,
which can be represented in the following compact form
\be
a^{(1)}_{s,0}(Q^2) = \frac{1}{L_0},~~
a^{(2)}_{s,1}(Q^2) =
a^{(1)}_{s,1}(Q^2) +
\delta^{(2)}_{s,1}(Q^2)
\,,
\label{as}
\ee
where
\be
L_i=\ln t_i,~~t_i=\frac{1}{z_i}=\frac{Q^2}{\Lambda_i^2},~~
\delta^{(2)}_{s,1}(Q^2) = - \frac{b_1\ln L_1}{L_1^2} ,~~
\label{ds}
\ee

As shown in Eqs.~(\ref{as}) and (\ref{ds}), in any PT order, the coupling $\ar(Q^2)$ contains its dimensional transmutation parameter
$\Lambda$, which is related to the normalization of $\alpha_s(M_Z^2)$,
where $\alpha_s(M_Z)=0.1176$ in PDG20 \cite{PDG20}.

{\bf $f$-dependence of the coupling $\ar(Q^2)$.}~~
The coefficients $\beta_i$ in (\ref{beta_i}) depend on the number $f$ of active quarks
which change the coupling $\ar(Q^2)$ at thresholds $Q^2_f \sim m^2_f$, where an additional quark enters the game for $Q^2 > Q^2_f$.
Here $m_f$ is the $\overline{MS}$ mass of the $f$ quark, e.g.,
$m_b=4.18 \pm 0.03$
GeV and $m_c=1.27 \pm 0.02$ GeV from PDG20 \cite{PDG20}.
\footnote{Strictly speaking, the quark masses in the $\overline{MS}$ scheme depend on $Q^2$ and $m_f=m_f(Q^2=m_f^2)$. The $Q^2$-dependence is rather slow and
  will not be discussed here.}
Thus, the coupling $a_s$ depends on $f$, and this $f$-dependence can be taken into account in $\Lambda$, i.e., it is $\Lambda^f$ that enters the above
Eqs.~(\ref{RenGro}) and (\ref{as}).

Relationships between $\Lambda_{i}^{f}$ and $\Lambda_{i}^{f-1}$, i.e.,
the so-called matching conditions
between $a_s(f,Q_f^2)$ and $a_s(f-1,Q_f^2)$
are known up to the four-loop order \cite{Chetyrkin:2005ia} in the $\overline{MS}$ scheme and are usually used
for $Q_f^2=m_f^2$, where these relations have the simplest form (see e.g.~\cite{Enterria} for a review).

Here we will not consider the $f$-dependence of $\Lambda_{i}^{f}$ and $a_s(f,M_Z^2)$, since we mainly consider the range of small $Q^2$ values and therefore use
$\Lambda_{i}^{f=3}$.

\section{Fractional derivatives}

Following \cite{Cvetic:2006mk},
we introduce the derivatives (in the $i$-th order of PT)
\be
\tilde{a}^{(i)}_{n+1}(Q^2)=\frac{(-1)^n}{n!} \, \frac{d^n a^{(i)}_s(Q^2)}{(dL)^n} \,,
\label{tan+1}
\ee
which are very convenient in the case of analytic QCD (see, e.g., \cite{Kotikov:2022JETP}).

The series of derivatives $\tilde{a}_{n}(Q^2)$ can successfully replace the corresponding series of $\ar$-powers. Indeed, each
derivative reduces the $\ar$-power, but is accompanied by an additional $\beta$-function $\sim \ar^2$.
Thus, each application of a derivative yields an additional $\ar$, and thus it is indeed possible to use series of derivatives instead of
series of $\ar$-powers.

At LO, the series of derivatives $\tilde{a}_{n}(Q^2)$ are exactly the same as $\ar^{n}$. Beyond LO, the relationship between $\tilde{a}_{n}(Q^2)$
and $\ar^{n}$ was established in \cite{Cvetic:2006mk,Cvetic:2010di} and extended to fractional cases, where $n$ is a non-integer $\nu$,
in Ref.~\cite{GCAK}.

Now consider the $1/L$-expansion of $\tilde{a}^{(k)}_{\nu}(Q^2)$. We can raise the $\nu$-th power of the result (\ref{as})
and then
restore $\tilde{a}^{(k)}_{ \nu}(Q^2)$ using the relations between $\tilde{a}_{\nu}$ and $\ar^{\nu}$ obtained in \cite{GCAK}
(see
Appendix B of \cite{Kotikov:2022sos}). 
Here we present only the final results, which have the form 
\footnote{
  Eq. (\ref{hR_i}) is similar to those used in \cite{BMS1,Bakulev:2006ex} for the expansion of
  ${\bigl({a}^{(i+1)} _{s,i}(Q^2)\bigr)}^ {\nu}$ in terms of  $a^{(1)}_{s,i}(Q^2)$-powers.}:
\bea
\z\tilde{a}^{(1)}_{\nu,0}(Q^2)={\bigl(a^{(1)}_{s,0}(Q^2)\bigr)}^{\nu} = \frac{1}{L_0^{\nu}},~
\tilde{a}^{(2)}_{\nu,1}(Q^2)=\tilde{a}^{(1)}_{\nu,1}(Q^2) +
\nu\,\tilde{\delta}^{(2)}_{\nu,1}(Q^2),~~\nonumber\\
\z\tilde{\delta}^{(2)}_{\nu,1}(Q^2)=
\hat{R}_1 \, \frac{1}{L_i^{\nu+1}},~~
\hat{R}_1=b_1 \Bigl[\Psi(1+\nu)+\gamma_{\rm E}
  + \frac{d}{d\nu}\Bigr]
\,,
\label{hR_i}
\eea
where
$\Psi(1+\nu)$ is the Euler $\Psi$-function and $\gamma_{\rm E}$ is Euler's constant.

The form
(\ref{hR_i}) of the $\tilde{\delta}^{(2)}_{\nu,1}(Q^2)$ corrections as the $\hat{R}_1$-operator is very important and allows us
to similarly present high-order results for the ($1/L$-expansion) of the analytic coupling.

\section{MA coupling}

We first show the LO results, and then go beyond LO, following our results (\ref{hR_i}) for the
usual strong coupling obtained in the previous section.

{\bf LO.}~~
The LO MA coupling $A^{(1)}_{{\rm MA},\nu,0}$
has the following form \cite{BMS1}
\be
A^{(1)}_{{\rm MA},\nu,0}(Q^2) = {\left( a^{(1)}_{\nu,0}(Q^2)\right)}^{\nu} - \frac{{\rm Li}_{1-\nu}(z_0)}{\Gamma(\nu)}=
\frac{1}{L_0^{\nu}}- \frac{{\rm Li}_{1-\nu}(z_0)}{\Gamma(\nu)} \equiv \frac{1}{L_0^{\nu}}-\Delta^{(1)}_{\nu,0}\,,
\label{tAMAnu}
\ee
where
\be
   {\rm Li}_{\nu}(z)=\sum_{m=1}^{\infty} \, \frac{z^m}{m^{\nu}}=  \frac{z}{\Gamma(\nu)} \int_0^{\infty}
\frac{ dt \; t^{\nu -1} }{(e^t - z)}
   \label{Linu}
\ee
is the Polylogarithm.
For $\nu=1$ we recover the famous Shirkov-Solovtsov results \cite{ShS}:
\be
\hspace{-0.5cm} A^{(1)}_{\rm MA,0}(Q^2) \equiv A^{(1)}_{\rm MA,\nu=1,0}(Q^2)
=\frac{1}{L_0}- \frac{z_0}{1-z_0}
\, .
\label{tAM1}
\ee

{\bf Beyond LO.}
Following Eq.~(\ref{tAMAnu})
for the LO analytic coupling,
we consider 
the derivatives of the
MA coupling, as
\be
\tilde{A}_{{\rm MA},n+1}(Q^2)=\frac{(-1)^n}{
  n!} \, \frac{d^n A_{\rm MA}(Q^2)}{(dL)^n}
\, .
\label{tanMA+1}
\ee

By analogy with the usual
coupling,
using the results (\ref{hR_i})
we have for the MA analytic coupling $\tilde{A}^{(2)}_{{\rm MA},\nu,1}$
the following expression:
\be
\tilde{A}^{(2)}_{{\rm MA},\nu,1}(Q^2) = \tilde{A}^{(1)}_{{\rm MA},\nu,1}(Q^2) +
\nu \,\tilde{\delta}^{(2)}_{{\rm A},\nu,1}(Q^2),
\label{tAiman}
\ee
where $\tilde{A}^{(1)}_{{\rm MA},\nu,1}$
is given in (\ref{tAMAnu}) and
\be
\tilde{\delta}^{(2)}_{{\rm A},\nu,1}(Q^2)= \tilde{\delta}^{(2)}_{\nu,1}(Q^2) -  \hat{R}_1 \left( \frac{{\rm Li}_{-\nu}(z_i)}{\Gamma(\nu+1)}\right)
\,.
\label{tdAman}
\ee
and $\tilde{\delta}^{(2)}_{\nu,1}(Q^2)$ and $\hat{R}_1$
are given in 
(\ref{hR_i}).

The analytical results for the MA
coupling $\tilde{A}^{(2)}_{{\rm MA},\nu,1}$
can be found in Ref.~\cite{Kotikov:2022sos}.
Here we present only the results for the case $\nu=1$:
\be
\tilde{\delta}^{(2)}_{{\rm A},\nu=1,1}(Q^2)
= \tilde{\delta}^{(2)}_{\nu=1,1}(Q^2)-P_{1,1}(z_1),~~
P_{1,1}(z)=b_1\Bigl[\overline{\gamma}_{\rm E}
  {\rm Li}_{-1}(z)+{\rm Li}_{-1,1}(z)\Bigr]\,,
\label{Pkz}
\ee
with
\be
\overline{\gamma}_{\rm E}=\gamma_{\rm E}-1,~~ {\rm Li}_{n,m}(z)= \sum_{m=1} \, \frac{\ln^k m}{m^n},~~
    {\rm Li}_{-1}(z)= \frac{z}{(1-z)^2}
    \, .
\label{Lii.1}
\ee

\section{Pion-photon transition form factor}

We investigate the pion-photon transition form factor (TFF), which is used to study chiral symmetry,
the quark-mass ratio, the characteristics of the pseudo-scalar meson decay, etc.,
within the framework of analytic QCD.
At the twist-two level,
the pion-photon TFF with one real and one virtual photon
can be decomposed into two parts: the perturbatively calculable coefficient function
and the non-perturbative
twist-two pion light-cone distribution amplitude
\cite{Efremov:1979qk}.
The valence quark TFF part
$Q^{2} F_{\rm V}^{\pi \gamma}(Q^{2})$ 
provides the dominant contribution and here
we limit ourselves to considering only this term.

Using the complete next-to-next-to-leading-order
QCD corrections~\cite{Braun:2021grd},
the perturbative expansion of the valence twist-two TFF part $F^{\gamma \pi (\tau=2)}_\text{V}$
can be expressed as \cite{Zhou:2023ivj}
\be
F^{\gamma \pi(\tau=2)}_\text{V}\left(Q^2\right)
= F^{\gamma \pi(\tau=2)}_\text{V,n=0}\left(Q^2\right)+\hat{b}_2(Q_0^2)\, F^{\gamma \pi(\tau=2)}_\text{V,n=2}\left(Q^2\right) + \hat{b}_4(Q_0^2)\,
F^{\gamma \pi(\tau=2)}_\text{V,n=4}\left(Q^2\right)\,,
\label{TFF1a}
\ee
where
\bea
&&Q^2 F^{\gamma \pi}_\text{V,n=0}\left(Q^2\right)=r_0^{[0]}+r_1^{[0]}a_s(Q^2)+(r_2^{[0]}+\beta_0\overline{r}_2^{[0]})a^2_s(Q^2)+O(a^3_s)\,,\label{TFF_2}\\
&&Q^2 F^{\gamma \pi(\tau=2)}_\text{V,n}\left(Q^2\right)= a^{d_n+1}_s(Q^2)\left(r_1^{[n]}+(R_2^{[n]}+\beta_0\overline{R}_2^{[n]})a_s(Q^2)+O(a^2_s)\right)\,,\label{TFF_n_2a}
\eea
and
\be
\hat{b}_n(Q_0^2)=\frac{b_n(Q_0^2)}{a^{d_n}_s(Q_0^2)},~~
d_{n=0}=0,~~d_{n=2}=\frac{50}{81},~~d_{n=4}=\frac{364}{405}\,,
\label{ha_n}
\ee
with 
the Gegenbauer moments \cite{Zhong:2021epq}
\be
b_{2}(Q_{0})=0.206\pm0.038,~~b_{4}(Q_{0})=0.047\pm0.011,~~Q_{0}=1 \mbox{ GeV}\,.
\label{an}
\ee

In the framework of
analytic QCD, we have
\be
F^{\gamma \pi(\tau=2)}_\text{V,MA}\left(Q^2\right)
= F^{\gamma \pi(\tau=2)}_\text{V,MA,n=0}\left(Q^2\right)+\hat{b}_2(Q_0^2)\, F^{\gamma \pi (\tau=2)}_\text{V,MA,n=2}\left(Q^2\right)
+\hat{b}_4(Q_0^2)\, F^{\gamma \pi(\tau=2)}_\text{V,n=4}\left(Q^2\right)\,,
\label{TFF1A}
\ee
where 
\bea
&&Q^2 F^{\gamma \pi(\tau=2)}_\text{V,MA,n=0}\left(Q^2\right)=r_0^{[0]}+r_1^{[0]}A_1(Q^2)+(r_2^{[0]}+\beta_0\overline{r}_2^{[0]})\tilde{A}_2(Q^2)+O(\tilde{A}_{3})\,,\label{TFF_0_NNLO_A}\\
&&Q^2 F^{\gamma \pi(\tau=2)}_\text{V,MA,n}\left(Q^2\right)= r_1^{[n]}\tilde{A}_{d_n+1}(Q^2)+(R_2^{[n]}+\beta_0\overline{R}_2^{[n]})\tilde{A}_{d_n+2}(Q^2)+O(\tilde{A}_{d_n+3}) \,.\label{TFF_n_2b_A}
\eea

\begin{figure}[t]
\centering
\includegraphics[width=0.68\textwidth]{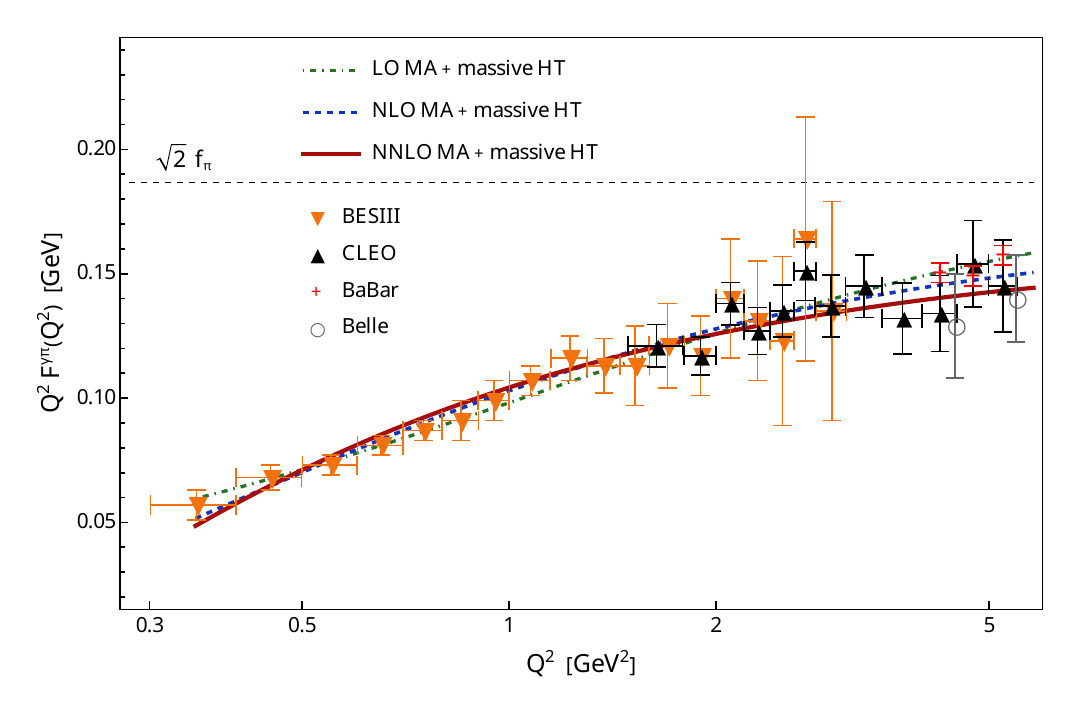}
\caption{\label{fig:as1352}
  The results (\ref{TFF1A})-(\ref{TFF_n_2b_A}) for the pion-photon TFF.
  Experimental data can be found in Refs.~\cite{Zhou:2023ivj} and \cite{Mikhailov:2021znq}.
}
\end{figure}

\vskip 0.3cm

By analogy with Refs.~\cite{Gabdrakhmanov:2023rjt,Gabdrakhmanov:2025vxw,Ayala:2017uzx}, where the Bjorken and Gross-Llewellyn Smith sum rules
were considered,
we perform a fit of experimental data for the pion-photon TFF
within the framework of
analytic QCD with the
``massive'' form \cite{Teryaev:2013qba} of the twist-four term.

We obtain a good agreement between the APT predictions and experimental data
(see Fig.~1).
\footnote{Similar good agreement was also observed for other values of $b_n(Q_0)$ (see \cite{Zemlyakov:2026rhe}).}
Moreover, the predictions
are in perfect agreement with the results of \cite{Mikhailov:2021znq},
obtained in the framework of the so-called method of light-cone sum rules in the form of dispersion relations \cite{Balitsky:1989ry} (see also \cite{Ayala:2018ifo}).

\section{Conclusions}

In this short paper, we have demonstrated the results obtained in our paper \cite{Kotikov:2022sos}.
It contains $1/L$-expansions of $\nu$-derivatives of the strong coupling $a_s$ expressed as
the $\hat{R}_m$ operators (see Eq. (\ref{hR_i}) for the operator $\hat{R}_1$) applied to the LO coupling $a_s^{(1)}$.
Using the same operators on $\nu$-derivatives of the LO MA coupling $A_{\rm MA}^{(1)}$,
various
representations were obtained for $\nu$-derivatives of the MA coupling,
i.e. $\tilde{A}_{\rm MA,\nu}^{(i)}$
in each $i$-th order of PT.
All results are presented in \cite{Kotikov:2022sos} up to the 5th order of PT,
where the corresponding QCD $\beta$-function coefficients are well known
(see \cite{Baikov:2016tgj}). In this paper, we have limited ourselves to the first two orders in order to avoid the most cumbersome
results.

High-order corrections are negligible in both asymptotics: $Q^2 \to 0$ and $Q^2 \to \infty$, and are nonzero in a neighborhood of the point $Q^2 =\Lambda^2$.
Thus, in fact, they represent only minor corrections to the LO coupling $A_{\rm MA}^{(1)}(Q^2)$.

An application of analytic QCD to the pion-photon TFF
is demonstrated in Section~5.
We obtain a good agreement between the QCD predictions and experimental data.\\

{\bf Acknowledgments}~
One of us (I.A.Z.) was supported in part by the Fellowship ANID Beca de Doctorado Nacional No.~21250067.
A.V.K.
was supported in part by the Russian Science Foundation grant No.~25-22-00576.
A.V.K.
thanks the Organizing Committee of the International conference "Particle Physics at Medium and High Energies"
(June 2-5, 2026, Protvino, Russia)
for the invitation.


\begin{thebibliography}{99}

\bibitem{Kotikov:2022sos}
A.~V.~Kotikov and I.~A.~Zemlyakov,
J. Phys. G \textbf{50}, no.1, 015001 (2023)

\bibitem{Bogolyubov:1959bfo}
N.~N.~Bogolyubov and D.~V.~Shirkov,
Intersci. Monogr. Phys. Astron. \textbf{3}, 1-720 (1959)

\bibitem{KoZe23}
A.~V.~Kotikov and I.~A.~Zemlyakov,
Phys. Rev. D \textbf{107}, no.9, 094034 (2023)

\bibitem{Baikov:2016tgj}
  P.A.~Baikov, K.G.~Chetyrkin,
  J.H.~K\"uhn,
Phys. Rev. Lett. \textbf{118}, no.8, 082002 (2017)

\bibitem{ShS}
D.V. Shirkov and I.L. Solovtsov,
Phys. Rev. Lett. \textbf{79}, 1209-1212 (1997);
D.V. Shirkov,
Theor. Math. Phys. \textbf{127}, 409-423 (2001);
Eur. Phys. J. C \textbf{22}, 331-340 (2001)

\bibitem{MSS}
K.~A.~Milton, I.~L.~Solovtsov and O.~P.~Solovtsova,
Phys. Lett. B \textbf{415}, 104-110 (1997)

\bibitem{BMS1}
A.~P.~Bakulev, S.~V.~Mikhailov and N.~G.~Stefanis,
Phys. Rev. D \textbf{72}, 074014 (2005)

\bibitem{Bakulev:2006ex}
A.~P.~Bakulev, S.~V.~Mikhailov and N.~G.~Stefanis,
Phys. Rev. D \textbf{75}, 056005 (2007);
JHEP \textbf{06}, 085 (2010)

\bibitem{Cvetic:2008bn}
G.~Cvetic and C.~Valenzuela,
Braz. J. Phys. \textbf{38}, 371-380 (2008)

\bibitem{Bogolyubov:1959vck}
N.~N.~Bogolyubov, A.~A.~Logunov and D.~V.~Shirkov,
Sov.Phys.JETP 10 (1960) 3, 574-581;
P.~J.~Redmond, Phys. Rev. \textbf{112}, 1404 (1958).

\bibitem{Bakulev:2008td}
A.~P.~Bakulev,
Phys. Part. Nucl. \textbf{40}, 715-756 (2009);
N.~G.~Stefanis,
Phys. Part. Nucl. \textbf{44}, 494-509 (2013)

\bibitem{Nesterenko:2003xb}
A.~V.~Nesterenko,
Int. J. Mod. Phys. A \textbf{18}, 5475-5520 (2003);
A.~V.~Nesterenko and J.~Papavassiliou,
Phys. Rev. D \textbf{71}, 016009 (2005)

\bibitem{PDG20}
Particle Data Group collaboration, P.A. Zyla et al., Review of Particle Physics, PTEP {\bf 2020}, 083C01 (2020).

\bibitem{Chetyrkin:2005ia}
K.~G.~Chetyrkin, J.~H.~Kuhn and C.~Sturm,
Nucl. Phys. B \textbf{744}, 121-135 (2006);
Y.~Schroder and M.~Steinhauser,
JHEP \textbf{01}, 051;
B.~A.~Kniehl, A.~V.~Kotikov, A.~I.~Onishchenko and O.~L.~Veretin,
Phys. Rev. Lett. \textbf{97}, 042001 (2006)


\bibitem{Enterria}
  D.~d'Enterria
  \textit{et al.},
J. Phys. G \textbf{51}, no.9, 090501 (2024)


\bibitem{Cvetic:2006mk}
  G.~Cvetic,
  C.~Valenzuela,
J. Phys. G \textbf{32}, L27 (2006);
Phys. Rev. D \textbf{74}, 114030 (2006)

\bibitem{Kotikov:2022JETP}
A.~V.~Kotikov and I.~A.~Zemlyakov,
JETP Lett. \textbf{115}, no.10, 565–56 (2022)

\bibitem{Cvetic:2010di}
G.~Cvetic, R.~Kogerler and C.~Valenzuela,
Phys. Rev. D \textbf{82}, 114004 (2010)

\bibitem{GCAK}
G.~Cveti\v{c} and A.~V.~Kotikov,
J. Phys. G \textbf{39}, 065005 (2012)

\bibitem{Efremov:1979qk}
A.~V.~Efremov and A.~V.~Radyushkin,
Phys. Lett. B \textbf{94}, 245 (1980);
G.~P.~Lepage and S.~J.~Brodsky,
Phys. Lett. B \textbf{87}, 359 (1979);
Phys. Rev. D \textbf{22}, 2157 (1980).

\bibitem{Braun:2021grd}
  V.~M.~Braun \textit{et al.},
Phys. Rev. D \textbf{104}, 094007 (2021);
J.~Gao \textit{et al.},
Phys. Rev. Lett. \textbf{128}, 6 (2022).


\bibitem{Zhou:2023ivj}
H.~Zhou, J.~Yan, Q.~Yu and X.~G.~Wu,
Phys. Rev. D \textbf{108}, no.7, 074020 (2023)

\bibitem{Zhong:2021epq}
  T.~Zhong \textit{et al.},
Phys. Rev. D \textbf{104}, 016021 (2021).

\bibitem{Gabdrakhmanov:2023rjt}
  I.~R.~Gabdrakhmanov \textit{et al.},
    JETP Lett. \textbf{118}, no. 7, 478-482 (2023);
Int. J. Mod. Phys. A \textbf{40}, no.04, 2450175 (2025)

\bibitem{Gabdrakhmanov:2025vxw}
  I.~R.~Gabdrakhmanov \textit{et al.},
Phys. Rev. D \textbf{113}, no.1, 014041 (2026)

\bibitem{Ayala:2017uzx}
  C.~Ayala \textit{et al.},
Int. J. Mod. Phys. A \textbf{33}, no.18n19, 1850112 (2018);
Eur. Phys. J. C \textbf{78}, no.12, 1002 (2018)

\bibitem{Teryaev:2013qba}
O.~Teryaev,
Nucl. Phys. B Proc. Suppl. \textbf{245}, 195-198 (2013);
V.~L.~Khandramai, O.~V.~Teryaev,
I.~R.~Gabdrakhmanov,
J. Phys. Conf. Ser. \textbf{678}, no.1, 012018 (2016)

\bibitem{Mikhailov:2021znq}
S.~V.~Mikhailov, A.~V.~Pimikov and N.~G.~Stefanis,
Phys. Rev. D \textbf{103}, no.9, 096003 (2021);
EPJ Web Conf. \textbf{258}, 03003 (2022)

\bibitem{Zemlyakov:2026rhe}
I.~A.~Zemlyakov, I.~L.~Chuev and A.~V.~Kotikov,
[arXiv:2608.00564 [hep-ph]].

\bibitem{Balitsky:1989ry}
I.~I.~Balitsky, V.~M.~Braun and A.~V.~Kolesnichenko,
Nucl. Phys. B \textbf{312}, 509-550 (1989);
A.~Khodjamirian,
Eur. Phys. J. C \textbf{6}, 477-484 (1999)

\bibitem{Ayala:2018ifo}
  C.~Ayala \textit{et al.},
Phys. Rev. D \textbf{98}, no.9, 096017 (2018);
EPJ Web Conf. \textbf{222}, 03017 (2019)

\end{thebibliography}
\end{document}